\documentclass[reprint, superscriptaddress, amsmath,amssymb, aps, prresearch, longbibliography, floatfix]{revtex4-2}

\usepackage{amsmath}
\usepackage{graphicx}
\usepackage{bm}
\usepackage{natbib}
\usepackage{dsfont}
\usepackage{epsfig}
\usepackage{feynmf}
\usepackage{blindtext, rotating}
\usepackage{mathtools}
\usepackage{dsfont}
\usepackage{subcaption}
\usepackage{physics}
\usepackage{amsfonts}
\usepackage{ragged2e}
\usepackage{siunitx}
\usepackage{comment}
\usepackage{soul}
\usepackage{lipsum}
\usepackage{hyperref} 
\hypersetup{breaklinks=true, colorlinks=true, citecolor=blue, linkcolor=cyan, urlcolor=blue,filecolor=blue}

\usepackage{xcolor} 

\DeclareCaptionJustification{justified}{\justifying}

\begin{document}

\title{Evolving Quantum Circuits}
\author{Daniel Tandeitnik}
\email{tandeitnik@aluno.puc-rio.br}
\affiliation{Department of Physics, Pontifical Catholic University of Rio de Janeiro, Rio de Janeiro 22451-900, Brazil}
\author{Thiago Guerreiro}
\email{barbosa@puc-rio.br}
\affiliation{Department of Physics, Pontifical Catholic University of Rio de Janeiro, Rio de Janeiro 22451-900, Brazil}

\begin{abstract}
We develop genetic algorithms for searching quantum circuits, in particular stabilizer quantum error correction codes. Quantum codes equivalent to notable examples such as the 5-qubit perfect code, Shor's code, and the 7-qubit color code are evolved out of initially random quantum circuits. We anticipate evolution as a promising tool in the NISQ era, with applications such as the search for novel topological ordered states, quantum compiling, and hardware optimization.
\end{abstract}


\maketitle

\section{Introduction}\label{sec:introduction}

Natural selection is a unifying idea in biology \cite{Bialek2012}. Through evolution and competition, systems can navigate the complexity landscape \cite{Kauffman2019} from small-sized molecules and molecule sets \cite{Xavier2020} to complex molecular machines \cite{Lau2017} to living organisms \cite{Alberts}. Through natural selection, complex designs such as eyes and brains can come into existence in the universe \cite{Kauffman2021}, and artificial evolution has been employed in the laboratory to produce molecules with desired optical response properties \cite{Piatkevich2018}. In view of that, an interesting engineering question is whether one can exploit evolution as a design tool for complex technology where human intuition and the standard deductive method might encounter difficulties or perhaps even fail. 

A natural field in which human intuition often encounters difficulties is quantum information science. Devising innovative means of producing complex quantum states and algorithms \cite{Shor2003} outside the scope of known quantum information primitives \cite{Martyn2021} is a challenging task \cite{NielsenChuang2016}, especially under the limitations imposed by present-day quantum hardware \cite{bravyi2018}. This motivates the main question of this work:
\textit{Can artificial selection be employed as an effective tool to design useful quantum circuits?}

Computer-assisted searches for new physical phenomena \cite{Iten2020} and laws of nature \cite{Cranmer2020, Cranmer2020b, Leary2021} comprise a very timely research topic, with notable examples including the search for new quantum optics experiments \cite{Krenn2016, Arrazola2019, Krenn2020}, resourceful states in quantum metrology \cite{Knott2016, Nichols2019, Rambhatla2020}, ground states in condensed matter systems \cite{Carleo2017}, the study of nonlocality \cite{Gao2018}, entanglement and Bell inequalities \cite{Canabarro2019, Krivachy2020} and the automated discovery of autonomous quantum error correction systems \cite{Wang2022} . Closely related to these developments, tools from artificial intelligence, genetic algorithms and competition have also been employed in chemistry, on the search for new pathways to organic molecules and closed autocatalytic reaction sets \cite{Wolos2020}, the efficient training of neural networks for image classification \cite{Real2019} and the evolution of deep learning algorithms through competition \cite{Olsson2018}. 
Here, we propose harvesting some of these tools and ideas within the context of quantum computing and quantum algorithms. 

Hilbert space is large \cite{Poulin2011, Susskind2018, Brandao2021} and -- similarly to the vast and complex landscape of the biosphere -- evolution may offer an efficient path to navigate its complexity. To test this idea, we apply genetic algorithms (GA) to the search for quantum circuits. As much as these ideas could benefit from a full scale quantum computer, there are not many such devices readily available for use at the present time \cite{Arute2019, Wu2021, Chow2021, Postler2021}. We therefore focus on stabilizer circuits \cite{Gottesman1997}, which are efficiently simulable on classical computers \cite{Aaronson2004, gidney2021}. The stabilizer formalism is the natural language of quantum error correction \cite{Gottesman1997}, leading us to evolve quantum error correction codes (QECCs) \cite{Roffe2019}. 

Within the framework of stabilizer QECCs, we will demonstrate that evolution, given the appropriate fitness landscape, can successfully produce known examples of error correcting codes, notably the \textit{Perfect} 5-qubit code \cite{Laflamme1996}, \textit{Shor's 9-qubit} code \cite{Shor1995}, the 7-qubit \textit{color} code \cite{calderbank1997quantum,kubica2018abcs} and novel examples of QECCs. These are arguably simple textbook examples but, as we will point out, the sample space (modulo equivalences) of circuits in which such codes live is so large that random search becomes prohibitive, thus proving the principle that evolution efficiently drives the search. 

Artificial selection might offer valuable opportunities in the current era of NISQ devices \cite{preskill2018} in which elementary quantum gates are costly and noisy. 
Generically, random stabilizer circuits are capable of generating good codewords for QECCs \cite{li2021}, but evolution can go beyond typicality in guiding the search for \textit{simple}, low-depth circuits more amenable to noisy devices.
Device specificity can also be taken into account by devising fitness landscapes in terms of the complexity geometry metric \cite{nielsen2005}, which penalizes gates according to hardware characteristics \cite{brown2019complexity}.

At present, our artificial selection algorithm can be compared to the evolution of simple bacteria in controlled laboratory conditions where the fitness landscape is simple and well understood \cite{Dekel2005}. With more complex fitness functions and the addition of quantum hardware we anticipate improvements and systematic means of devising complex quantum circuits in various applications beyond stabilizer circuits and QECCs including but not limited to learning unitaries \cite{kiani2020}, quantum compiling \cite{heyfron2018, nam2018, jones2022}, and hardware specific tailor-made circuits \cite{brown2019complexity, Arute2019}.
We highlight that while evolution may be complementary to known circuit optimisation schemes \cite{nam2018}, its main strength relies on the possibility of creating novel and creative quantum circuits. All scripts used in this work are available in the GitHub repository \cite{Tandeitnik_Evolving_Quantum_Circuits_2022}.

This paper is organized as follows. In Section \ref{GA} we introduce the GA applied to quantum circuits. Section \ref{toy} is dedicated to a simple application of the GA, demonstrating its capabilities within a well-understood context. Next, we apply the tools of evolution to the search for QECCs in Section \ref{qeccs}. We conclude with a brief discussion and outlook.

\section{Genetic Algorithm}\label{GA}

For about 4.28 billion years \cite{dodd2017evidence}, millions of living species go through a continuous optimization process, what Darwin termed the evolution of species \cite{darwin2004origin}. The evolutionary process occurs through the interaction of populations of species with their habitat, whose function is to select individuals with a greater aptitude to grow, thrive, and reproduce. Mathematically, one can allude to the environment as a fitness function whose domain and codomain are individuals of a population and a measurement of how well an individual is adapted to its habitat, respectively. The fitness function then becomes a cost function of the natural selection optimization problem.

A key point of evolution is its robustness to noise, an often missing feature in standard optimization methods \cite{fogel1994introduction}. The environment is not static: through natural unpredictable events, it transforms in a chaotic fashion leading to an ever-changing fitness landscape. Inspired by the evident success of life in thriving through evolution in dynamic, noisy environments, researchers have been using algorithms simulating natural selection to solve mathematical optimization problems as early as 1957 \cite{fraser1957simulation, barker1958simulation,bremermann1962optimization,bremermann1968numerical,bremermann1964evolution,reed1967simulation,sampson1976adaptation}, broadly termed evolutionary algorithms \cite{bartz2014evolutionary}. 

In this section we make a brief introduction to the biological evolutionary process emphasizing its genetic point of view. Next, GAs to evolve Clifford quantum circuits are introduced.

\subsection{The evolutionary process}\label{sec:gen_algorithm}

Given a population of individuals belonging to the same species, one may divide its evolution process into four stages \cite{fogel1994introduction}: reproduction, mutation, competition, and selection. In broad terms, reproduction is the transmission of genetic material from parents to their offspring. Mutations are minute stochastic errors that occur during the transmission of genes. Competition and selection drift the population towards individuals better adapted to the environment, where the fittest individuals reproduce passing along their genes to future generations, while the unfitted perish removing their genes from the gene pool. An evolutionary algorithm aims at emulating, to some extent, this cycle for a population of potential solutions to an optimization problem until a threshold is reached.

The evolution process hinges on the representation of individuals as their genetic material, commonly referred to as the \textit{genotype}, and a screening method that favors certain genotypes over others. Further, the genotype is a collection of DNA strands made of elementary building blocks from a finite set of possibilities, namely the four nucleotides adenine, thymine, guanine, and cytosine \cite{roth2019genomic}. Considering sexual reproduction, the genotype design naturally leads to the \textit{genetic operations} of \textit{crossover} and \textit{mutation}. Crossover is a recombination of the parent's DNA, where whole segments of DNA are interpolated to form the offspring's DNA at conception. Crossover promotes the dispersion of good DNA blocks among the population, as natural selection gradually removes bad blocks, and mutation enables the search for new solutions not reachable through recombination.

Given the genetic operators, we have only an aimless, random walk through the genotype sample space due to their stochastic nature. Natural selection is responsible for drifting the walk in the direction of genotypes that beget fitter individuals. Natural selection results from the interplay between the environment's restrictions on the individuals and the \textit{phenotype} each one displays. The phenotype is the genetic expression of the genotype, i.e., the set of physical attributes displayed by the individual \cite{pierce2012genetics} --- this set determines the individual's probability of reproduction.

\begin{figure}
    \includegraphics[width=0.35\textwidth]{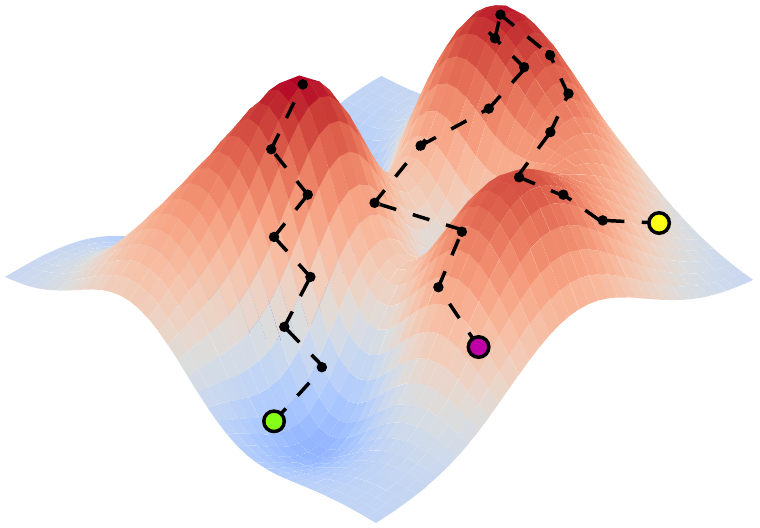}
    \caption{Illustration of three distinct populations climbing a hypothetical fitness landscape. At an initial time, each population starts at some low point of the landscape (colored circles) and stochastically rises towards the peaks. The mutation rate regulates the size of each step. The height of the surface represents the reproduction rate of a given genotype.}
    \label{fig:fitness_landscape}
\end{figure}

A helpful way of visualizing the evolutionary optimization process is with a fitness landscape \cite{gavrilets2004fitness,kauffman1987towards}. The fitness landscape is a mathematical construct that maps genotypes to reproduction rates. The environment defines a function in which the domain is the space of all possible genotypes and whose codomain is the reproduction rate. Distances on the landscape are defined as the closeness of two genotypes, i.e., how similar are the phenotypes they spawn. Figure \ref{fig:fitness_landscape} exemplifies a representation of an arbitrary fitness landscape. The rationale is that, at an initial time, the population genotype set is centered at some point in the landscape. As the generations go by, the population stochastically drifts towards the surface peaks due to natural selection and the genetic operators. Naturally, the fitness landscape in Figure \ref{fig:fitness_landscape} is a schematic representation, as actually producing such a graph may be impossible due to the complex nature of the problem at hand.

\subsection{Genetic algorithms applied to Clifford circuits}

We now describe how we mimic nature's evolutionary process in a GA applied to Clifford quantum circuits. We note three main aspects we need to cover to build a GA that simulates nature: (1) a genetic representation of the tentative solutions, (2) a method to implement the genetic operators of crossover and mutation, (3) a selection mechanism that distinguishes between solutions regarding the optimization problem.

\begin{figure}
    \centering
    \includegraphics[width=0.3\textwidth]{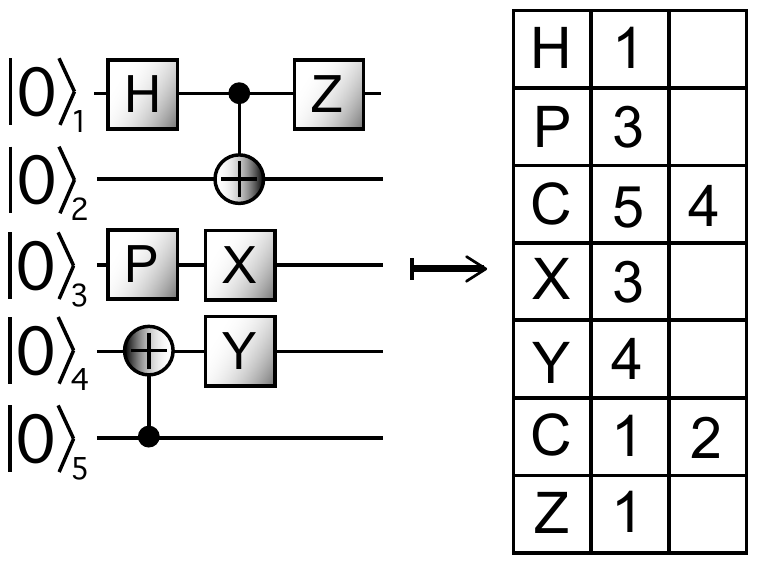}
    \caption{A random quantum circuit and its genotype. Each row of the genotype is a gate in ascending order (from top to bottom) of application, with the columns storing the operator and the indices of the target qubits. The $\rm{CNOT}$ is abbreviated as $\rm{C}$.}
    \label{fig:genetic_representation}
\end{figure}

To illustrate how we genetically represent quantum circuits in this work, consider the random circuit depicted in Figure \ref{fig:genetic_representation}(a) and its genetic expression in Figure \ref{fig:genetic_representation}(b). We genetically represent a circuit composed of $t$ gates as a $t\times 3$ array whose rows are gates in ascending order of application. The first column stores the operator, and the second and third the indices of the affected qubits. If the gate is a $\rm{CNOT}$, the second (third) column is the control (target) qubit index. For single-qubit gates the third column is ignored.

\begin{figure}
    \centering
    \includegraphics{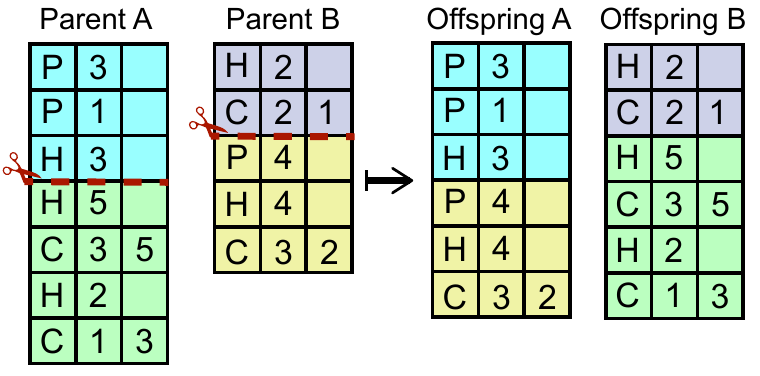}
    \caption{Example of a crossover between two genotypes. The parents' genotypes are divided at randomly chosen points and their offspring are built by stacking the pieces.}
    \label{fig:circuit_crossover}
\end{figure}

Crossovers and mutations naturally become row operations by expressing quantum circuits as a matrix. There is a freedom of choice in defining how each genetic operator works. The most appropriate option often arises from experimentation since the efficiency of a GA is strongly dependent on the optimization problem itself \cite{bartz2014evolutionary}. For instance, the crossover can be performed at one-point or at multiple points \cite{fogel1994introduction,bartz2014evolutionary,holland1992adaptation}. Additionally, we may use more than two parents at the reproduction step, thus having more than two genotypes to crossover \cite{eiben1994genetic,ting2005mean}. For simplicity, we worked solely with one-point crossovers between the genotypes of two parents. Hence, given two arbitrary parents A and B, their genotypes are divided at split points randomly selected according to a uniform probability distribution. Offspring A(B) is formed by stacking the top portion of parent A(B) on top of the bottom portion of parent B(A). Figure \ref{fig:circuit_crossover} shows an example of the crossover operation.

Mutations happen to the offspring's genotype after a crossover procedure. With equal probabilities, it is chosen whether (a) the mutation modifies an existing gate or (b) it inserts a new gate into the circuit. Then,
\begin{itemize}
        \item  If (a): a random row of the genotype is uniformly selected to be altered. The row contents are overwritten by a new gate uniformly selected between $\lbrace I, H, P, \rm{CNOT}\rbrace$. If the identity is picked, the entire row is deleted;
        \item If (b): a random insertion point on the genotype is uniformly selected. A new row with a new uniformly selected gate chosen between $\lbrace H, P, \rm{CNOT}\rbrace$ is inserted at the chosen point. 
\end{itemize}

\begin{figure}
    \centering
    \includegraphics[width=0.3\textwidth]{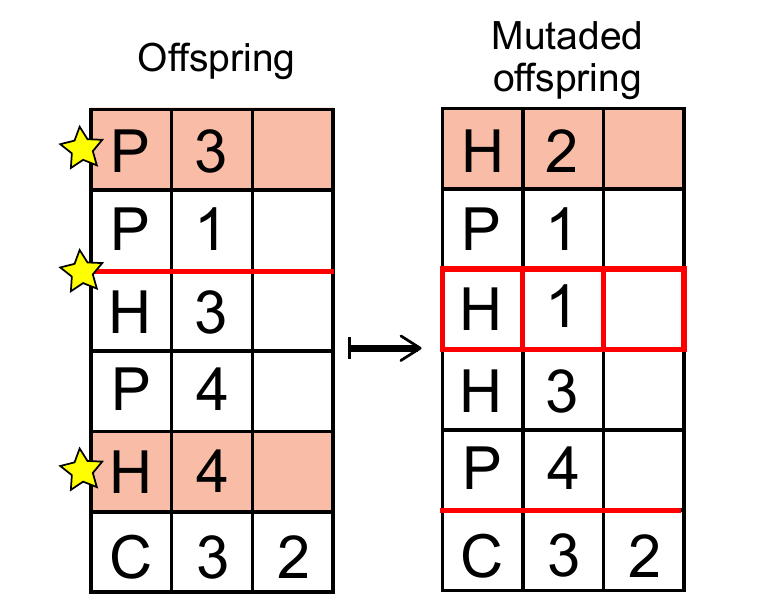}
    \caption{Illustration of the three types of mutation that may occur to a circuit. From top to bottom: the first gate is replaced by a Hadamard gate on qubit 2, a new row in inserted between the second and third rows, and an identity replaces the fifth gate, hence deleting it.}
    \label{fig:circuit_mutation}
\end{figure}

\noindent Figure \ref{fig:circuit_mutation} shows an example displaying the three kinds of mutations that can happen to an offspring genotype.

Just as in nature, we also consider that each genotype gives rise to a phenotype in GAs. We regard the phenotype of a given circuit as a set of real-valued numbers that quantify its properties. For instance, we extensively consider the depth (the length of the longest path from the beginning of a circuit to its end) as an essential parameter since it is a measure of how long it takes to execute a circuit. Thus, from a population of circuits, one can sort the individuals by their depth and use it as a selection bias, e.g., considering circuits with lower depth as better suited. Hence, the \textit{selection mechanism} for quantum circuits works in the following way: 

\begin{enumerate}
    \item The phenotype of each individual is evaluated, where the optimization problem determines the list of assessed parameters;
    \item Each individual's fitness is evaluated via a fitness function $\mathcal{F}$ whose arguments are the phenotypes. By convention, we define $\mathcal{F}$ such that higher fitness individuals are considered better solutions;
    \item For breeding selection, a probability of reproduction is associated to each individual proportional to its relative fitness with respect to the rest of the population. A roulette wheel selection system \cite{fogel1994introduction} is employed to pick two individuals to mate, i.e., to go through the process of crossover/mutation.
\end{enumerate}

\noindent The most computationally expensive part of the GA is calculating the phenotype. Moreover, coming up with a proper set of phenotype parameters might be challenging. Clear understanding of the optimization problem is hence key in devising the defining parameters which capture an optimal solution, as well as methods to calculate these phenotypes.

Finally, after the size of the population reaches an established maximum limit -- recall reproduction adds two new individuals at each iteration -- we may purge the worst circuits. We regard this process as the competition aspect of the GA since it emulates the limited growth of biological populations and the death of ill-adapted individuals.

\begin{figure}[ht]
    \centering
    \includegraphics{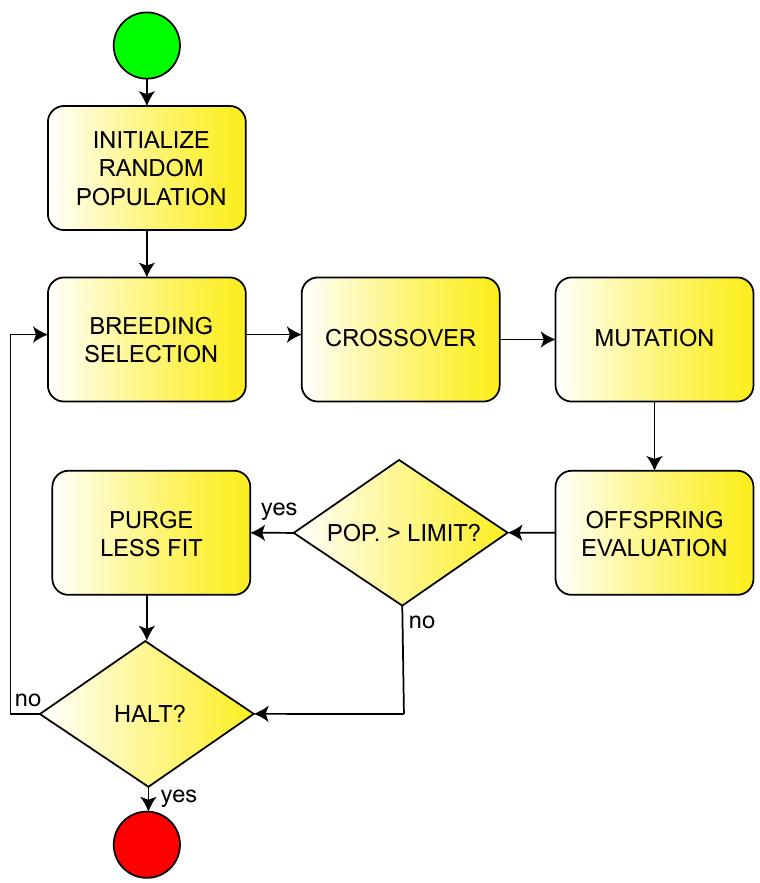}
    \caption{Decision tree for the genetic algorithm. The halt decision gate evaluates when a termination condition is met.}
    \label{fig:GA_decision_tree}
\end{figure}

With all the fundamental elements of the GA defined, we build the basic scheme of the simulated evolutionary cycle. At the start, an initial population of random circuits is initialized with each individual's fitness evaluated and stored. The GA then works via the iteration of a cycle of selective breeding, crossover, mutation, and population purge until a termination criterion is met. Termination criteria can be a maximum number of iterations or a target fitness value. Each cycle marks a \textit{generation}. Putting it all together, Figure \ref{fig:GA_decision_tree} displays the complete decision tree of the GA.

\section{Toy model}\label{toy}

As a first test of the capability of the GA as a searching tool for quantum circuits we introduce a simple toy problem.
A circuit $ U $ acts on a 1D lattice of qubits, which we now label by the indices $x \in \lbrace 1, ..., n\rbrace $, initially in the state $\ket{\psi_0} = \ket{0}^{\otimes n}$. The resulting state after application of the circuit is $\ket{\psi} = U \ket{\psi_{0}}$.  Define the density matrix $ \rho_{x} $ as the state obtained by splitting the chain at $x$ and tracing out all qubits to the left of $x$ in the final state $ \ket{\psi} $. The von Neumann entropy of $ \rho_{x} $ is denoted $S(x)$, and we define the \textit{mean entropy} generated by a circuit as
\begin{equation}
    \langle S \rangle = \dfrac{1}{n} \sum_{x} S(x).
\end{equation}

Consider then the following problem: \textit{find quantum circuits that generate the largest possible $ \langle S \rangle $ with the least possible circuit depth $ D $.} We propose the fitness function
\begin{equation}\label{eq:toy_fitness}
    \mathcal{F} = \langle S \rangle / D \ .
\end{equation}

\noindent Circuits that maximize \eqref{eq:toy_fitness} solve the problem. Note that solutions can be found by deductive reasoning as follows. Subadditivity of the von Neumann entropy implies that $S(x)$ can change by at most one from one qubit to the next \cite{Nahum2017},
\begin{equation}
   \vert S(x+1) - S(x) \vert \leq 1.
\end{equation}

\noindent The maximum mean entropy of a circuit is therefore given by
\begin{equation}
    \langle S \rangle_{\mathrm{max}} = \dfrac{1}{n} \left( \sum_{M =1}^{n/2} M + \sum_{M=1}^{n/2 - 1} M \right) = \dfrac{n}{4}.
    \label{eq:max_entropy}
\end{equation} 

\begin{figure}[h] 
    \centering
    \includegraphics[width=0.3\textwidth]{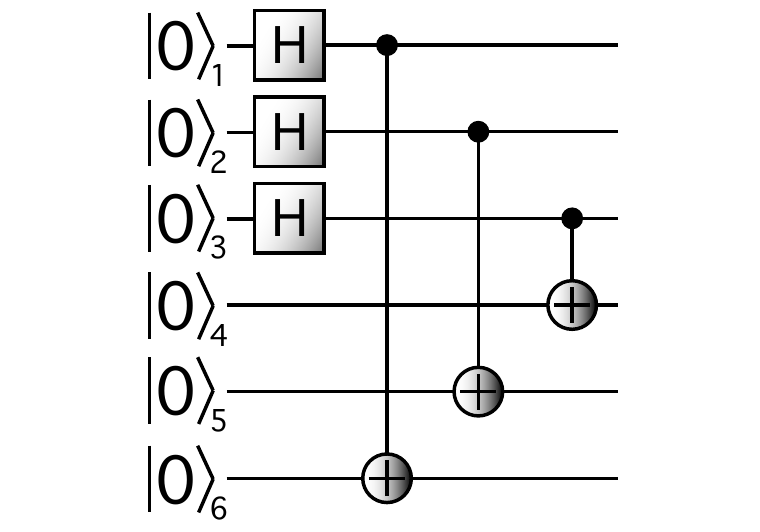}
    \caption{A $6$-qubit circuit solving the toy problem. Permutations of the $\rm{CNOT}$s along the time direction, as well as of the target qubits yields the same solution.}\label{fig:toy_sol_example}
\end{figure}

\noindent The minimum value of depth for a circuit generating non-vanishing entropy is $D_{\mathrm{min}} = 2$. Hence, the highest possible value for $\mathcal{F} $ is $\mathcal{F}_{\mathrm{max}} = n/ 8$. There are multiple solutions that maximize $ \mathcal{F} $, one example of which is shown in Figure \ref{fig:toy_sol_example} for $ n = 6 $ qubits. Note that the circuit produces a tensor product of Bell pairs organized in a specific way within the 1D lattice. 

\begin{figure} 
    \centering
    \includegraphics{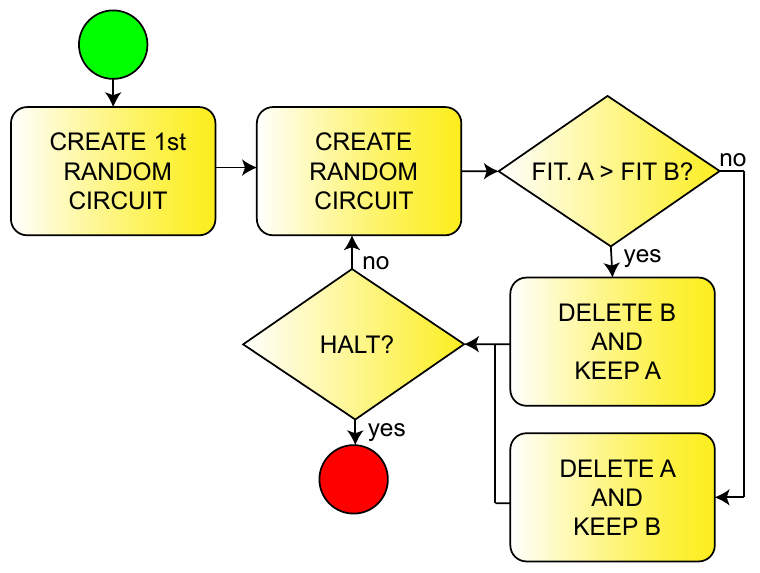}
    \caption{Decision tree for the RS. The algorithm generates random circuits and saves the best circuit ever generated until a termination condition -- maximum number of generations or maximum fitness value -- is reached. }\label{fig:RS_decision_tree}
\end{figure}

To determine whether a GA provides advantage in the search for the solution, we compare its performance to a purely random search (RS). The decision tree used for RS is shown in Figure \ref{fig:RS_decision_tree}. Since the bottleneck of both algorithms is the fitness evaluation, their time performance per iteration is nearly identical, i.e., a generation for each method takes almost the same time to execute. Hence, we chose to compare the methods by how fast each can converge to a solution within a given number of generations.

\begin{figure}[ht]
    \centering
    \includegraphics{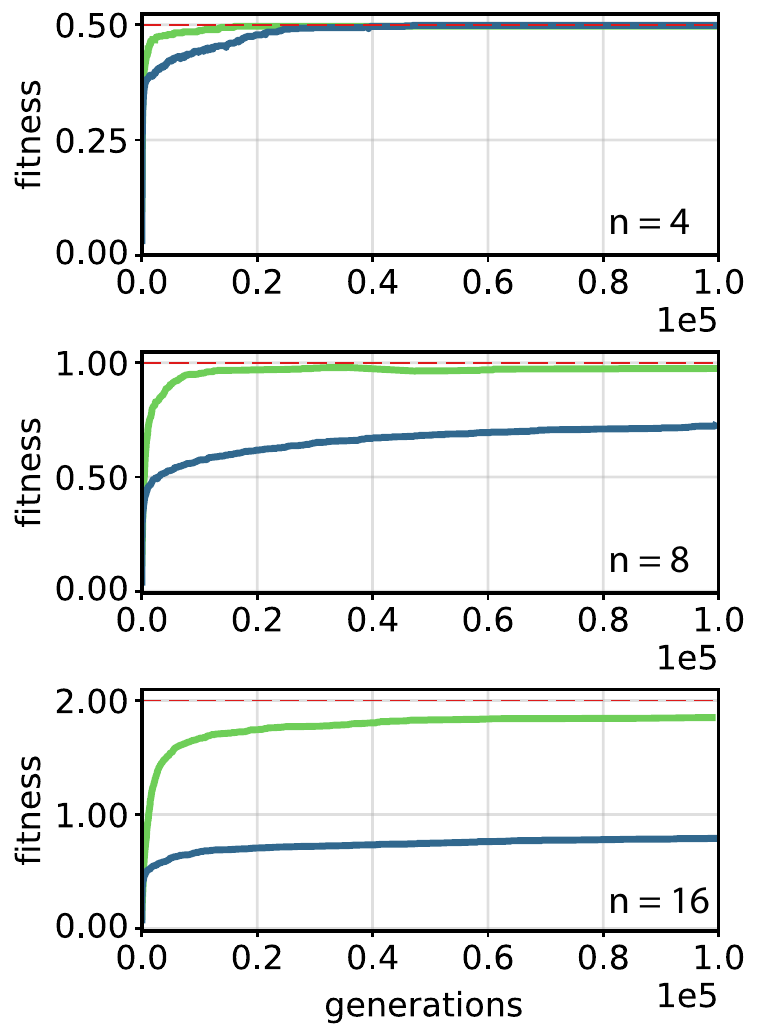}
    \caption{Evolutionary search using a genetic algorithm (green line) versus random search (blue line) for $n=4, 8, 16$ qubits in the search for a circuit maximizing Eq. \eqref{eq:toy_fitness}. Each curve is the average over $100$ runs. The dashed red line represents the maximum fitness given by $ n /8 $; see Eq. \eqref{eq:max_entropy} and the main text.}\label{fig:toymodel_result}
\end{figure}

The total number of solutions to the problem is very small in comparison to all possible $n$-qubit circuits composed of $t$ gates. We expect further that the ratio of solutions to possible circuits decreases significantly with the number of qubits $n$. If the GA is to have an advantage over RS, we should expect this edge to grow as we increase $n$. Figure \ref{fig:toymodel_result} shows the fitness values of the best circuit in the population as a function of generation number for three cases with increasing number of qubits given by  $n = 4, 8$ and $ 16 $. We note that each curve is the average over $100$ runs, and that over all runs the optimal circuits are generated many times by the GA. For $ n = 4$ we can see that RS is able to find solutions. Nevertheless, the GA has an increasing advantage over RS as we increase the number of qubits, as expected. This can be seen by the widening gap between the GA and RS traces in Figures \ref{fig:toymodel_result}(a), (b) and (c). The red dashed lines show the maximum attainable fitness values, for reference.
\begin{figure}[h] 
    \centering
    \includegraphics{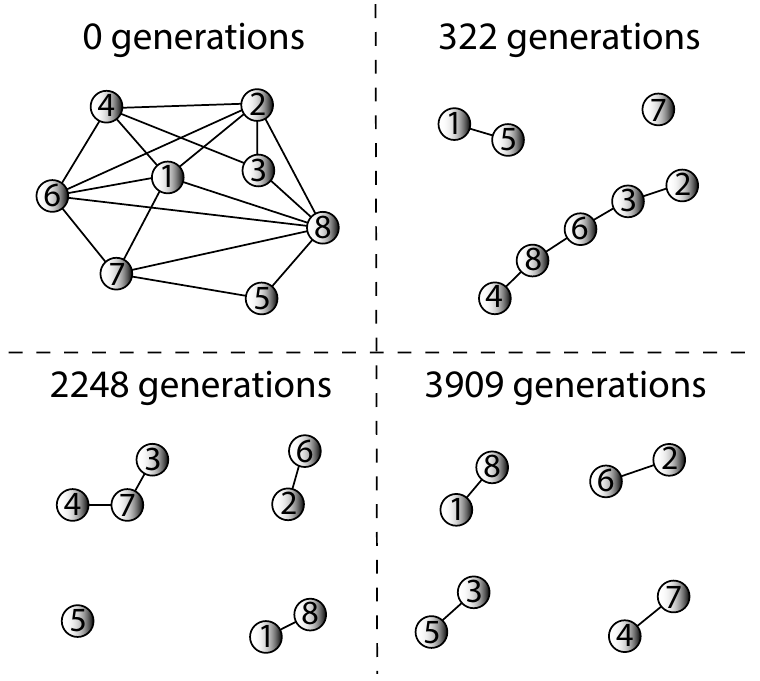}
    \caption{Evolution of the topology of the fittest circuit in an evolutionary search simulation for $n =8$ qubits. The edges represent the qubits indexed by lattice position, and the vertices correspond to CNOT gates. }\label{fig:evo_topo}
\end{figure}

We can visualize the topology of a quantum circuit as a graph, where each node corresponds to a qubit and vertices represent qubit interactions, implemented here as $\rm{CNOT}$ gates. This graph representation highlights the necessary hardware architecture a quantum computer needs to have to execute the circuit, i.e. which qubits need to be coupled. Figure \ref{fig:evo_topo} shows the typical example of an initial randomly generated circuit topology, and its subsequent evolution towards the topology of the optimal solution, as depicted in Figure \ref{fig:toy_sol_example} corresponding to Bell pairings of the qubits. We see that starting from complicated random circuits artificial selection can perform a directed search towards high fitness individuals in a much shorter time than RS.

\section{Evolutionary search for QECCs}\label{qeccs}

Given the appropriate fitness function, can a GA evolve quantum error correction codes such as the 5-qubit \textit{perfect} code \cite{Laflamme1996}, the 7-qubit \textit{color} code \cite{calderbank1997quantum,kubica2018abcs}, and \textit{Shor's 9-qubit} code \cite{Shor1995}? These are examples of stabilizer codes, meaning they are generated by Clifford gates. As in the toy example, efficient simulation in classical computers is therefore granted by the Gottesman-Kill theorem \cite{Gottesman1997}.

Defining the appropriate fitness function that will efficiently direct the search for the desired QECCs is crucial, and not a straightforward task. The crux of the matter involves defining a phenotype set that captures the expected features of a good error correction code. For the toy example in the previous section, the definition of the problem itself contained the relevant phenotype leading to the form of the fitness function. Backed by the content developed in Appendix \ref{sec:appendix_qecc}, we now present metrics capable of measuring the effectiveness of a circuit in correcting quantum errors and a method for evaluating such metrics.

\subsection{QECC fitness function}

One can represent a given QECC by a set of two or more mutually orthogonal codewords $\lbrace\ket{c_i}\rbrace$. The codewords define a group of syndrome operators $\lbrace S_i\rbrace$ employed to detect and correct errors up to weight $t$ following the scheme shown in Figure \ref{fig:generic_stabilizer_qecc} of Appendix \ref{sec:appendix_qecc}. With the syndrome operators and a subset of errors $\mathcal{E}$ we require our code to detect and correct, one builds the syndrome table. The table can then be used to decide if the set of codewords forms a functional QECC by checking if there are undetectable and uncorrectable errors. We reiterate how undetectable and uncorrectable errors are classifieds:

\begin{itemize}
    \item Syndromes represented by bit strings of $0$-s correspond to undetectable errors. Thus, given the syndrome table, the number of undetectable errors is obtained by counting how many errors return $0$-stringed syndromes;
    \item If more than one error is associated with distinctive syndromes, all 2-on-2 combinations are cross-checked with the shared stabilizers of all codewords. If one combination fails, we classify the errors associated to the syndrome as uncorrectable.
\end{itemize}

\noindent Let $e_{\mathrm{und}}$ and $e_{\mathrm{unc}}$ be the number of undetectable and uncorrectable errors associated to a set of codewords, respectively. Then, define the \textit{corrigibility degree} $ \mathcal{C} $ as
\begin{equation}
    \mathcal{C} \equiv (\vert \mathcal{E} \vert-e_{\mathrm{und}}-e_{\mathrm{unc}})/\vert \mathcal{E} \vert
\end{equation}

\begin{figure}[h]
    \centering
    \includegraphics{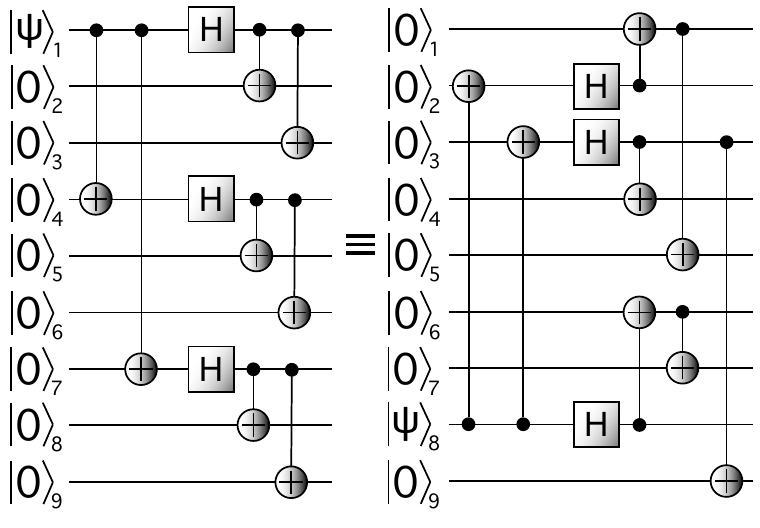}
    \caption{Standard Shor's $[[9,1,3]]$ EC and an example of an equivalent circuit.}\label{fig:shor_equiv_circ}
\end{figure}

We use the corrigibility degree as our main phenotype to evolve QECCs, however its evaluation assumes the possession of a tentative set of codewords. 
To employ the GA, we must relate codewords to Clifford circuits, as we now explain.
Consider Shor's code as an illustrative example. The code can be represented by its codewords or by an encoding circuit (EC) as depicted in Figure \ref{fig:shor_equiv_circ}(a).
Note, however, the EC in itself is insufficient to determine the codewords; one must determine which initial states are acted upon by the EC. 
For the particular circuit portrayed in Figure \ref{fig:shor_equiv_circ}(a), the $\ket{\psi}$ ket in the first register implicitly informs that the initial states are taken to be $\lbrace\ket{00\dots0} $ or $\ket{10\dots0}\rbrace$. 
Furthermore, different circuits may generate equivalent codewords for Shor's code in the sense that both sets of codewords form a $[[9,1,3]]$ QECC. For example, consider the circuit shown in Figure \ref{fig:shor_equiv_circ}(b): if it acts on the initial states $\lbrace\ket{0\dots000},\ket{0\dots010}\rbrace$ it effectively generates equivalent codewords.

The main takeaway is that if we are going to evolve ECs, we need an algorithm capable of producing different sets of tentative codewords from various combinations of initial states. For instance, if the GA evolved the circuit from Figure \ref{fig:shor_equiv_circ}(b) and only tried to use $\lbrace\ket{00\dots0},\ket{10\dots0}\rbrace$ as the initial states to form the codewords, it would erroneously conclude that the circuit does not constitute a good QECC.

In Appendix \ref{sec:app_algorithm}, we describe the procedure to evaluate different sets of tentative codewords given a Clifford circuit regarded as an EC for a QECC. Given this procedure, we are ready to introduce the fitness function. 

Consider a tentative EC with circuit depth $D$. Let $V_{\perp}$ be the set of all codewords $\lbrace \ket{c_0},\ket{c_i}\rbrace$ with associated corrigibility degree $\mathcal{C}_{i}$. The fitness associated to the EC is then given by 
\begin{equation}\label{eq:fitness_qecc1}
\mathcal{F} = \max_{ V_{\perp} } \ F
\end{equation}
\noindent where
\begin{equation}\label{eq:fitness_qecc2}
    F =  w \times \mathcal{C}_{i} - D.
\end{equation}
where $ w $ is a predefined weight.

Given a specific problem, there is generally no unique definition for the fitness function. Specifically in the search for QECCs, the form \eqref{eq:fitness_qecc2} is the result of intuitive reasoning and empirical tinkering. For instance, we tested and excluded functional forms containing terms proportional to $ \mathcal{C}/D$ as we observed these to enforce a tendency of evolving low depths in detriment of corrigibility $ \mathcal{C}$, while we imperatively wish the highest possible value for $ \mathcal{C}$. 
A simple solution is to give greater weight to phenotypes that maximize $ \mathcal{C}$, implemented by multiplying it by a factor of $w$. We find $ w = 1000 $ to yield satisfactory results.  

\subsection{Results}

\begin{figure*}[t] 
   \centering
   \includegraphics{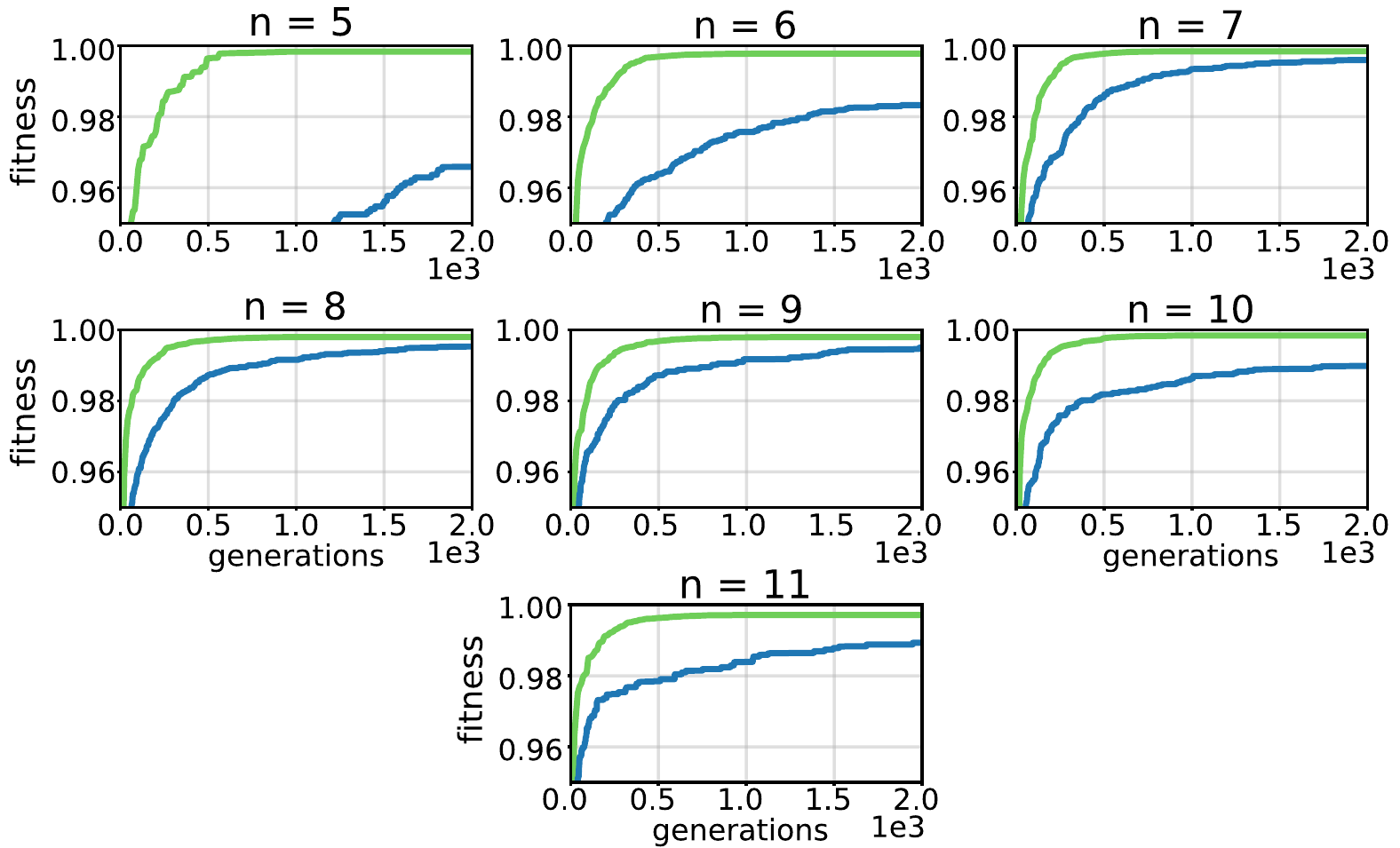}
    \caption{Evolutionary search using the genetic algorithm (green line) versus random search (blue line) for QECCs with $n=5, 6, 7, 8, 9, 10, 11$ qubits. Each curve is the average over $100$ runs of the population's best fitness. The fitness values are normalized.}\label{fig:qecc_evo_plot}
\end{figure*}

We have applied the GA to the problem of searching QECCs capable of correcting up to single-qubit errors using the fitness function as defined in Eqs. \eqref{eq:fitness_qecc1} and \eqref{eq:fitness_qecc2}. We tested the GA performance against RS for a range of qubit overheads, from $n = 5$ to $n = 11$. Figure~\ref{fig:qecc_evo_plot} displays the results. In all seven cases, the GA showed a clear advantage over RS by being able to evolve functional, low-depth QECCs in just a few hundred generations. 
Indeed, the GA was capable of finding multiple examples of circuits equivalent -- in terms of distance -- to the perfect
5-qubit and Shor’s 9-qubit code starting from random circuits and with no built-in specific insights referring to the quantum codes.


\begin{figure}[h] 
    \centering
    \includegraphics{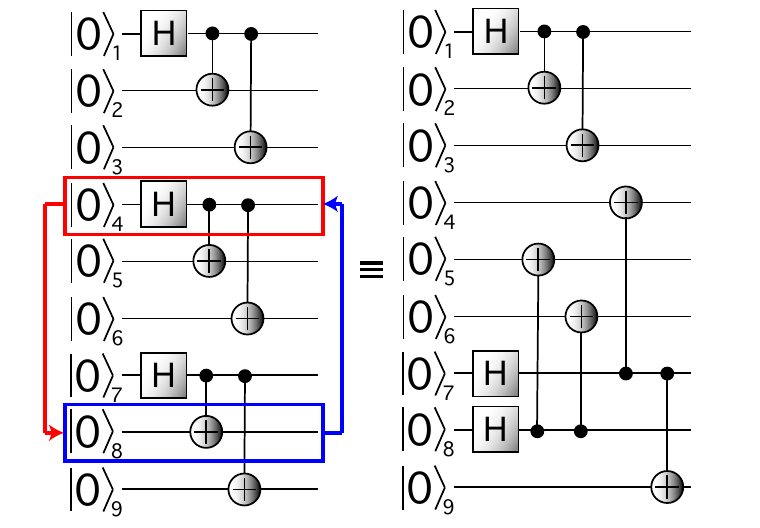}
    \caption{Generation of an equivalent EC for Shor's code by permutation of two qubit registers. Any permutation of the register produces equivalent ECs. }\label{fig:shor_equiv_permut}
\end{figure}

An unanticipated result was that finding QECCs either by GA or RS seems to become easier as the number of physical qubits increases, up until $n = 8$. Recall that the number of pure stabilizer states scales as $O(2^{n^2})$ \cite{Aaronson2004}. We therefore expected it would be easier to find codes with fewer physical qubits, while plots in Figure \ref{fig:qecc_evo_plot} show the opposite behavior from $n = 5$ to $8$.
We conjecture that the number of possible QECCs rapidly grows with $n$, making it easier to find codes for $n$ between $5$ to $8$. For instance, inequality \eqref{eq:overhead_t1}, demonstrated in Appendix \ref{sec:qubit_overhead}, indicates that the number of single-qubit errors grows linearly. In contrast, the number of potential syndrome codes grows exponentially with $n$, thus allowing a wider range of codes to exist. Moreover, the issue of equivalent ECs enlarges the number of solutions in the sample space. Figure \ref{fig:shor_equiv_permut} illustrates this point where any register permutation creates a new equivalent circuit. Therefore, the freedom to exchange registers makes a single solution appear $n!$ times. Given these considerations, it is conceivable that for $n$ equals 5 to 8, the number of solutions grows faster than the sample space of solutions.

Nevertheless, while RS indeed found solutions within the given limit of $2,000$ generations, it rarely found low depth examples. In contrast, the GA consistently evolved QECCs with depth lower than $6$, as shown in Table \ref{tab:table_results_qecc}. Even though it appears from the graphs in Figure \ref{fig:qecc_evo_plot} that RS becomes more efficient as $n$ increases, the data in Table \ref{tab:table_results_qecc} suggests otherwise.  The GA had a success rate above $70\%$ in all cases in finding QECCs with $D \le 6$. On the other hand, RS varied between $4\%$ and $52\%$ success rates showing an abrupt downward trend starting from $n = 9$. Finally, the GA was able to evolve circuits with depth as low as $4$, while RS was not capable. We expect that for larger values of $n$, RS becomes increasingly inefficient. Furthermore, the difference in the speed at which each method converges to a solution is substantial. Note the speed of convergence becomes increasingly important for large $n$ since computational costs scales accordingly.

\begin{table}
\center
\caption{
Percentage of solutions found by the GA and RS with depth less than or equal to 6, 5 and 4 within the maximum limit of 2,000 generations for qubit overhead $n$.}
\label{tab:table_results_qecc}
\center
\hskip-1.5cm\begin{tabular}{c|ccc|ccc}
\hline
\hline 
 \, & \multicolumn{3}{c|}{GA}  &  \multicolumn{3}{c}{RS}\\
 $n$ & $D \le 6$  &  $D \le 5$  &  $D \le 4$  &  $D \le 6$  &  $D \le 5$  &  $D \le 4$\\
\hline
5 & $81\%$ & $48\%$ & $19\%$ & $12\%$ & $2\%$ & $0\%$ \\
6 & $70\%$ & $27\%$ & $6\%$ & $15\%$ & $2\%$ & $0\%$ \\
7 & $93\%$ & $82\%$ & $56\%$ & $52\%$ & $9\%$ & $0\%$ \\
8 & $92\%$ & $72\%$ & $33\%$ & $46\%$ & $5\%$ & $0\%$ \\
9 & $90\%$ & $70\%$ & $32\%$ & $38\%$ & $2\%$ & $0\%$ \\
10 & $82\%$ & $58\%$ & $14\%$ & $4\%$ & $0\%$ & $0\%$ \\
11 & $72\%$ & $44\%$ & $12\%$ & $9\%$ & $0\%$ & $0\%$ \\
\hline \hline
\end{tabular}
\end{table}

To further demonstrate the applicability of GAs in finding specific codes, we modified the fitness landscape to specialize in the search for color codes \cite{kubica2018abcs}. Color codes are a particular class of error correction codes contained in the broader family of \textit{topological} QECCs \cite{kubica2018abcs,kitaev2003fault}. An important feature of topological codes is their modularity, i.e., the main code is assembled by patching elementary repeated pieces. Modularity enables the scalability of code circuits with each module requiring only nearest-neighbor interactions, relieving a major hardware constraint. These, including other reasons, place topological codes among the leading prospects for actual hardware implementation \cite{nickerson2014freely, sete2016functional, o2016silicon}.

We set out to evolve color codes in two-dimensional (2D) lattice configurations with nearest-neighbor qubit interactions. As we will see, a simple modification of the previous fitness function suffices to guide evolution.
As a defining phenotype, we note 2D color codes are members of the CSS class \cite{gottesman1996class,calderbank1997quantum}. This means that, given a set of codewords, if their common stabilizer set can be generated by stabilizers constructed only from $X$s or $Z$s independently -- that is, each generator is made only by $X$s or $Z$s -- then the code is of the CSS type \cite{calderbank1996good,steane1996multiple,kubica2018abcs}. 
We can then build a metric quantifying how close to this requirement a code is, similarly to the corrigibility degree introduced previously. 

Given a set of codewords, we measure its CSS degree `$\mathcal{CSS}$' by constructing a generator set, for their joint stabilizers, with the maximum number of operators made only by $X$s or $Z$s possible. We define $\mathcal{CSS}_{i}$ of a tentative pair of codewords $ i $ as the ratio of operators that satisfies the CSS criterion by the total number of elements in the generator set. Hence, to drive the GA towards two-dimensional color codes, we modify Eq. \eqref{eq:fitness_qecc2} according to
\begin{equation}\label{eq:fitness_color}
    F =  w \times \mathcal{C}_{i} + w' \times \mathcal{CSS}_{i}-D.
\end{equation}
\noindent Again, the weight $w'$ is used to enforce codes that satisfy the CSS criterion. We use $ w' = 1000$.

\begin{figure}
    \centering
    \includegraphics[width=0.3\textwidth]{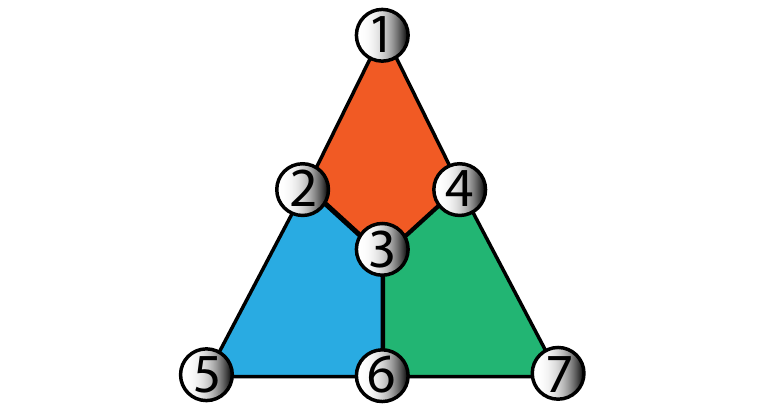}
    \caption{Lattice arrangement of the $7$-qubit color code.}\label{fig:color_lattice}
\end{figure}

\begin{figure}
    \centering
        \includegraphics{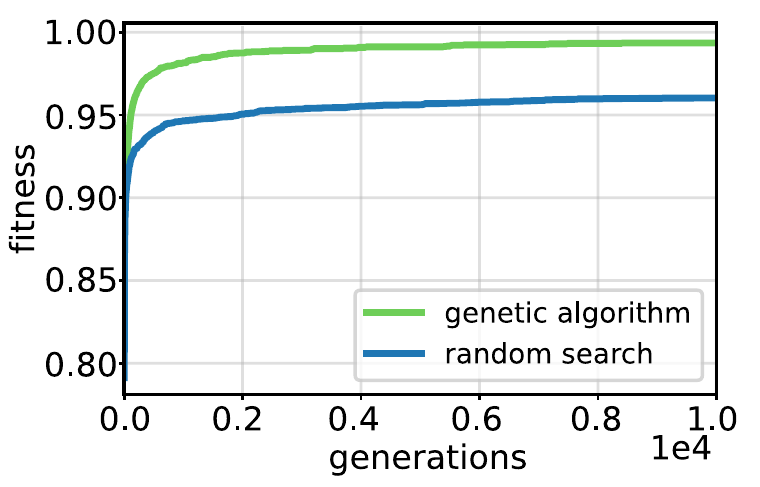}
    \caption{Evolutionary search using a genetic algorithm versus random search for the $7$-qubit color code. Each curve is the average over $100$ runs of the population's best fitness. The fitness values are normalized according to the standard fitness of the circuit from \cite{Postler2021} .}\label{fig:qecc_color}
\end{figure}

The simplest 2D color code is the triangular $7$-qubit code \cite{kubica2018abcs} with topology as illustrated in Figure \ref{fig:color_lattice} \cite{bermudez2017assessing}. While searching for color codes using the GA, we have used the 7-qubit code as a benchmark. Figure \ref{fig:qecc_color} shows the performance of the GA versus RS. The fitness values are normalized by a target-fitness for a color code circuit taken from \cite{Postler2021}. Noting the greater difficulty in evolving color codes, we have increased the generation limit to $10,000$. 
Despite the appearance that RS performs comparably well, in reality the GA had a success rate of $54\%$ while RS never found any color code.

As a final remark we highlight that calculating phenotypes can be a difficult task. For instance, it is generally expected that computing the code distance is exponentially hard in the number of physical qubits $ N $ \cite{gullans2020dynamical}. This difficulty can be overcome, for example, by noting that the error-correcting capacity of a code is related to the entanglement structure between regions in the code lattice quantified by the mutual information between the region and its complement \cite{li2021}. Computing the mutual information of partitions of a stabilizer state is polynomial in $ N $ \cite{Aaronson2004}, and thus a suitable fitness function for large numbers of qubits may be devised based on this method. 
This reinforces that despite evolution being blind, its proper application relies on mathematical knowledge and creativity in the fitness definition, especially in the limit of large systems.

\section{Conclusion}

In conclusion, we have employed genetic algorithms to the search for quantum error correction codes starting from an initially random population of quantum circuits.
Through the introduction of suitable fitness functions, the genetic algorithm was capable of repeatedly outperforming random search and evolving examples of QECCs equivalent to Shor's $9$-qubit code, as well as the $5$-qubit perfect code.
We have also employed the genetic algorithm in the search for topological codes, notably the 2D 7-qubit code, successfully demonstrating that the genetic method can be used in the targeted search for quantum circuits with specific properties, notably the evolution of topological order.

We anticipate at least three directions of future research in which ideas closely related to the ones introduced here might lead to interesting developments within the broader field of quantum information and computing.

First, evolution might offer valuable means to search for novel topological codes. It might be possible to encode topological features of quantum states into an appropriate fitness function thus enabling the automated search for new topologically ordered states characterized by distinct values of the topological entanglement entropy \cite{kitaev2006, levin2006}.

Second, quantum algorithm compilation \cite{harrow2001quantum} is a challenging problem that will become increasingly important as noisy intermediate-scale quantum (NISQ) computers emerge \cite{preskill2018,khatri2019quantum}. Hence, the development of quantum compiling methods is an active field \cite{khatri2019quantum, venturelli2018compiling, booth2018comparing, cincio2018learning, maslov2008quantum, booth2012quantum, chong2017programming, heyfron2018efficient, haner2018software}. In principle, GAs could provide a promising tool for the automation of quantum compiling and depth reduction. 

\begin{figure}[t!]
    \centering
    \includegraphics{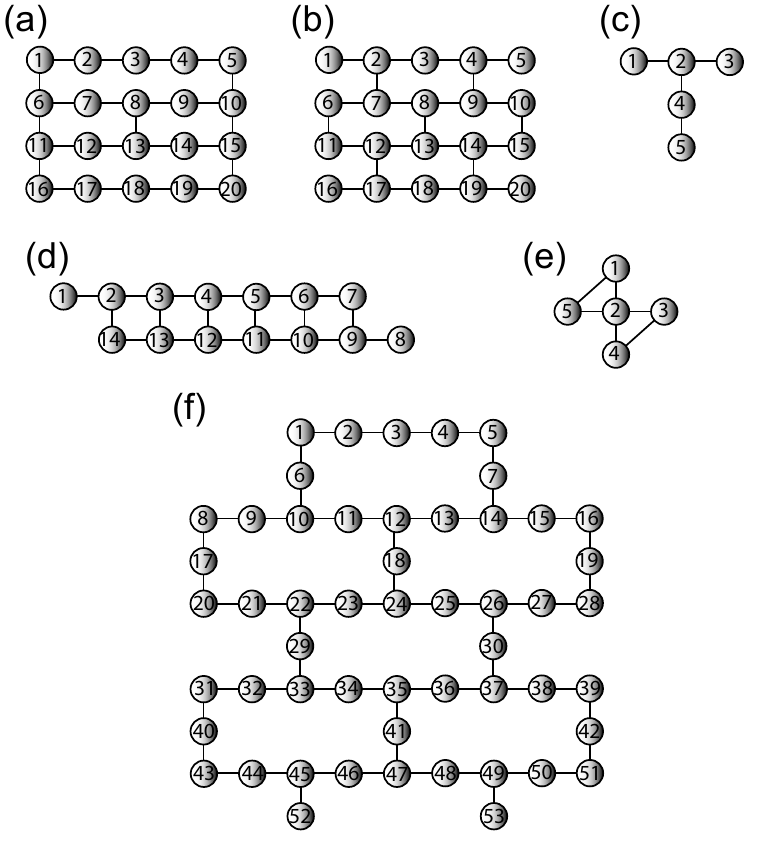}
    \caption{IBM quantum lattices, adapted from \cite{IBMlattices}: (a) $20$-qubits systems Johannesburg and Poughkeepsie; (b) $20$-qubits systems Almaden, Boeblingen, and Singapor; (c) $5$-qubits systems Ourense, Valencia, and Vigo; (d) $14$-qubits system Melbourne; (e) $5$-qubits system Yorktown; (f) $53$-qubits system Rochester. GAs can be used to devise tailor circuits for each specific hardware topology.}\label{fig:IBM_lattices}
\end{figure}

Third, one may translate quantum hardware specifications into metrics incorporated in the fitness function to produce tailor-made algorithms. As an illustration, Figure \ref{fig:IBM_lattices} shows the available lattices of the quantum system devices IBM offers on their Quantum service \cite{IBMlattices}. Note that each device geometry imposes interaction restrictions on two-qubit gates. Given a device with a particular lattice, we could incorporate penalty factors for circuits that disobey the topology restrictions of the quantum device under consideration. 
Additionally, some gates may be harder to implement due to details of the experimental implementation. Brown and Susskind \cite{brown2019complexity} introduce the concept of \textit{unitary complexity}, quantifying the difficulty in implementing a given unitary provided physical properties and constraints of the setup. The unitary complexity can also be used as a penalty factor in the fitness function, driving the search for simple circuits for a given hardware implementation.

Finally, we conclude by highlighting how remarkable it is that an unsupervised genetic algorithm with a few hundred lines of code can evolve in minutes celebrated results that required the insightful minds of Ray Laflamme, Peter Shor and Robert Calderbank to come about. 
As quantum computers improve in number of qubits and more complex topologies, genetic algorithms may yet prove to be an invaluable optimization method in the quantum computing engineer's toolbox.

\section*{Acknowledgements}
We thank Bruno Suassuna, Igor Brand\~ao, Lucianno Defaveri, George Svetlichny, Stuart Kauffman, Ernesto Galv\~ao and Guilherme Tempor\~ao for useful discussions. This study was financed in part by Conselho Nacional de Desenvolvimento Cient\'ifico e Tecnol\'ogico (CNPq), and by the Fundação de Amparo à Pesquisa do Estado do Rio de Janeiro (FAPERJ). This study was financed in part by the Coordenação de Aperfeiçoamento de Pessoal de Nível Superior - Brasil (CAPES) - Finance Code 001. We would like to thank the support received by CNPq Scholarship No. 132606/2020-8, FAPERJ Scholarship No. 2021.01394.9, and CNPq Scholarship No. 140197/2022-2.


\bibliography{main.bbl}

\newpage
\appendix

\section{Brief review of QECC stabilizer codes}\label{sec:appendix_qecc}
\begin{figure*}[t]
    \centering
    \includegraphics{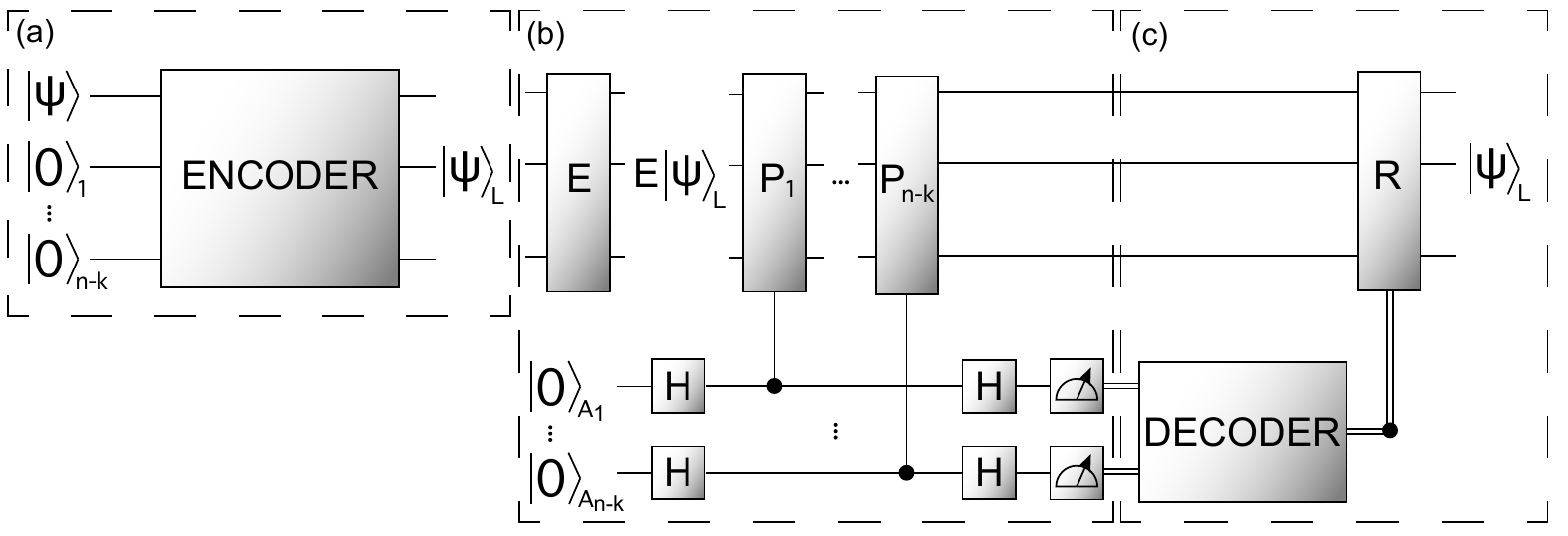}
    \caption{The generic circuit of a $[[n,k,d]]$ stabilizer error correction code. (a) A data register $\ket{\psi}_D$ is entangled with $n-k$ redundancy qubits via an EC to form the logical-state $\ket{\psi}_L$. (b) After a potential error $E$ occurs, ancilla qubits are attached to $\ket{\psi}_L$ and $m$ syndrome measurements $P_i$ are performed. The result of the measurements produces the syndrome. (c) With the syndrome, one queries the syndrome table, and the appropriate correction $R$ is appointed and applied. This process is represented by the Decoder gate. The double-line channels symbolizes classical communication.}
    \label{fig:generic_stabilizer_qecc}
\end{figure*}

The main objective of a QECC is to protect $k$ data-qubit registers from a set of errors $\mathcal{E}$. By protection, it is meant that the code must be able to detect and correct any error in $\mathcal{E}$. Figure \ref{fig:generic_stabilizer_qecc} summarizes the general structure of a $[[n,k,d]]$ stabilizer error correction code. It is divided into three stages: \textit{encoding}, \textit{syndrome extraction} and \textit{correction}. We now describe each stage in detail.

On the encoding stage, Figure \ref{fig:generic_stabilizer_qecc}(a), a $k$-qubit data-state $\ket{\psi}_D$ is entangled with $m = n-k$ auxiliary qubits via an EC forming a $n$-qubit logical state $\ket{\psi}_L$. The EC defines the set of codewords $\mathcal{C} = \lbrace \ket{c_i}_L\rbrace_i$ which specify how the other stages of the code function. In general, $\vert\mathcal{C}\vert = 2^k$. For example, for $k = 2$, the encoding stage maps $\lbrace \ket{00},\ket{01},\ket{10},\ket{11}\rbrace$ into four mutually orthogonal codewords in an expanded Hilbert space.

During syndrome extraction, errors are detected by performing $m$ syndrome measurements as shown in Figure \ref{fig:generic_stabilizer_qecc}(b). The codewords define a group of common stabilizers spanned by $m$ generators \cite{Gottesman1997}. Let $\mathcal{P} = \lbrace P_i \rbrace$ be a set of generators for the common stabilizer group of $\mathcal{C}$. We refer to the elements of $\mathcal{P}$ as syndrome operators. It follows that syndrome operators satisfy the following properties \cite{Joschka2019}:

\begin{enumerate}
    \item\label{en:S_prop_1} $\mathcal{P} \subseteq \mathcal{G}_n$;
    \item\label{en:S_prop_2} $P_i\ket{\psi}_{L,j} = +1\ket{\psi}_{L,j} \, \forall \,  i,j$;
    \item\label{en:S_prop_3} $[P_i,P_j] = 0 \, \forall \, i,j$,
\end{enumerate}

\noindent where $\mathcal{G}_n$ is the general $n$-qubit Pauli group defined as the set composed of all tensor product combinations of the elements of $\mathcal{G}_1 = \lbrace \pm I,\pm iI, \pm X,\pm iX, \pm Y,\pm iY, \pm Z,\pm iZ \rbrace$. Property \ref{en:S_prop_2} ensures that the syndrome measurements do not further disturb the damaged logical state, and property  \ref{en:S_prop_3} allows one to perform measurements in any order. 

Let $E \in \mathcal{E}$ be an error that occurred between the encoding and syndrome extraction stages. The effect of each syndrome measurement is to map $E\ket{\psi}_L$ into the superposition
\begin{align}\label{eq:syndrome_map}
    E&\ket{\psi}_L\ket{0}_{A_i} \rightarrow \nonumber\\ &\frac{1}{2}\left[(I+P_i)E\ket{\psi}_L\ket{0}_{A_i} + (I-P_i)E\ket{\psi}_L\ket{1}_{A_i}\right].
\end{align}

\noindent Note that $E$ necessarily either commutes or anti-commutes with $S_i$ since $E,P_i \in \mathcal{G}_n$. If $E$ and $P_i$ commutes (anti-commutes), the final state is unequivocally $E\ket{\psi}_L\ket{0}_{A_i}$ ($E\ket{\psi}_L\ket{1}_{A_i}$). Therefore, each syndrome measurement can be understood as a deterministic measurement of the state with the outcome revealing whether the error commutes or anti-commutes with the syndrome operator. At the end of the syndrome extraction stage, one is left with a binary syndrome string of length $m$ whose $i$-th entry encodes whether $P_i$ and $E$ commutes or not.

Given $\mathcal{E}$ and $\mathcal{P}$, one builds the so-called \textit{syndrome table} relating each error to the corresponding syndrome string it generates. As an illustration, Table \ref{tab:3_qubit_table} shows the syndrome table for the $3$-qubit code with $\mathcal{P} = \lbrace Z_1Z_2,Z_2Z_3\rbrace$, and considering errors with weight one on three qubits. Note that the syndromes for $Z_i$ are composed only of zeros, which is the same syndrome for $E = I$ since the identity commutes with any operator. These errors are classified as \textit{undetectable} \cite{Gottesman1997} as they are not distinguishable from $I$. 

\begin{table}[]

\centering
\caption{$3$-qubit code syndrome table for single-qubit errors.}
\label{tab:3_qubit_table}
\hskip-1.5cm\begin{tabular}{cc|cc}
\hline
\hline 
Error & Syndrome & Error & Syndrome\\
\hline
$X_1$ & $10$ & $Z_1$ & $00$\\
$X_2$ & $11$ & $Z_2$ & $00$\\
$X_3$ & $01$ & $Z_3$ & $00$\\
\hline\hline
\end{tabular}
\end{table}

Finally, in the correction stage one prescribes an operator $R$ for which
\begin{equation}
    RE\ket{\psi}_L = \ket{\psi}_L.
\end{equation}

\noindent Since Pauli operators square to the identity, $R$ is in principle identical to the appointed error guided by the syndrome table. The decoder gate of Figure \ref{fig:generic_stabilizer_qecc}(c) is a crosscheck, performed in a classical computer, between the extracted syndrome and its related error on the syndrome table. This scheme functions perfectly for \textit{non-degenerate codes}, where a one-to-one correspondence between errors and syndromes exists. On the other hand, for \textit{degenerate} codes multiple errors can produce the same syndrome. For a successful correction, all two-on-two combinations of errors with the same syndrome must stabilize all codewords. Consider an arbitrary correction code with codewords $\lbrace \ket{c_i}_L\rbrace$, and let $\lbrace E_i \rbrace$ be a set of errors with the same syndrome. One requires,
\begin{equation}\label{eq:degenerate_condition}
    E_iE_j\ket{c_k}_L = +\ket{c_k}_L \, \forall i,j,k
\end{equation}

\noindent for $\lbrace E_i \rbrace$ to be correctable. If the above condition is met, applying any element of $\lbrace E_i \rbrace$ will restore the logical state even though it is impossible to single out which error actually took place. If for some pairing $E_iE_j$ Equation \eqref{eq:degenerate_condition} is not satisfied, the set $\lbrace E_i \rbrace$ is classified as \textit{uncorrectable}, since it is impossible to decide the proper correction operation.

\section{Generating sets of codewords}\label{sec:app_algorithm}

Given an EC, applying it to the $\ket{0}^{\otimes n}$ state generates a first possible codeword $\ket{c_0}$. Recollecting that codewords form a set of mutually orthogonal states, we create a method to build $2^{n}-1$ mutually orthogonal states to $\ket{c_0}$ which will give us a set of $2^n$ potential codewords (since we work with qubits, a $n$-dimensional Hilbert space is spawned by $2^n$ states) $\lbrace \ket{c_i} \rbrace$. In possession of $\lbrace \ket{c_i} \rbrace$, it remains to evaluate the \textit{corrigibility degree} $ \mathcal{C} $ of subsets. To generate states orthogonal to $\ket{c_0}$, we find a set of logical $\mathcal{X} \equiv \lbrace\bar{X}_i\rbrace$ operators such that
\begin{align}
    \bar{X}_i\ket{c_0} &= \ket{c_i} \label{eq:x_condition_1}\\
    \bra{c_i}\ket{c_j} &= \delta_{ij}\label{eq:x_condition_2}
\end{align}

\noindent $\forall i \in \lbrace 1,\dots,2^{n-1}\rbrace$. 

There is a systematic method to build a particular (non-unique) set $\mathcal{X}$ that satisfies the above equations starting from the computational basis. Consider the $n$-qubit computational basis. Starting from $\ket{\psi_{0,I}} = \ket{0}^{\otimes n}\,$ (the first subscript $0$ refers to $\ket{0}^{\otimes n}$ and the $I$ subscript stands for the identity), it is straightforward to verify that the logical $\lbrace\bar{X}_{i,I}\rbrace$ operators  that take $\ket{\psi_{0,I}}$ to the other states of the basis $\ket{\psi_{i,I}}$ -- which are mutually orthogonal by definition -- are all the $2^n-1$ tensor product combinations of Pauli letters  $X$ and $I$ possessing at least one $X$. Define $[\bar{X}_I]$ as a $2^n-1 \times n$ matrix whose rows are $\lbrace\bar{X}_{i,I}\rbrace$:
\begin{equation}
    [\bar{X}_I] = \begin{bmatrix} 
        X & I   & I &\dots & I & I\\
        I & X   & I &\dots & I & I\\
        \vdots & \vdots & \vdots & \vdots & \vdots & \vdots \\
        I & I   & I &\dots & I & X\\
        X & X   & I &\dots & I & I\\
        \vdots & \vdots & \vdots & \vdots & \vdots & \vdots \\
        X & X   & X &\dots & X & X
    \end{bmatrix}.
\end{equation}

\noindent Notice that $\lbrace\bar{X}_{i,I}\rbrace$ satisfies Equations \eqref{eq:x_condition_1} and \eqref{eq:x_condition_2}. Let $\ket{\psi_{0,U}} = U\ket{\psi_{0,I}}$, where $U$ is some unitary computation. The operator $U$ transforms each $\bar{X}_{i,I}$ into $\bar{X}_{i,U} \equiv U\bar{X}_{i,I}U^\dagger$ such that
\begin{align}
    \bra{\psi_{0,U}}\bar{X}_{i,U}\ket{\psi_{0,U}} &= \bra{\psi_{0,I}}U^\dagger U\bar{X}_{i,I}U^\dagger U\ket{\psi_{0,I}} \nonumber\\
    &= \bra{\psi_{0,I}}\bar{X}_{i,I}\ket{\psi_{0,I}} \nonumber\\
    &= 0\label{eq:X_U}
\end{align}
\noindent $\forall i$. Each $\bar{X}_{i,U}$ is distinct since if $\bar{X}_{i,U} = \bar{X}_{j,U}$ for some $i,j$ , then
\begin{equation}
     U\left(\bar{X}_{i,I}-\bar{X}_{j,I}\right)U^\dagger = 0.
\end{equation}
\noindent Since $U \neq 0$ and $\lbrace\bar{X}_{i,I}\rbrace$ are different by construction, then forcibly $i = j$. We conclude that $\lbrace\bar{X}_{i,U}\rbrace$ forms a set of logical operators whose elements take $\ket{\psi_{0,U}}$ into $2^n-1$ unique mutually orthogonal states $\ket{\psi_{i,U}}$. Taking the particular case of $U$ as an EC, $\mathcal{X} = \lbrace\bar{X}_{i,\rm{EC}}\rbrace$ is the set we seek.


There is a significant computational cost in evaluating the \textit{corrigibility degree} $ \mathcal{C} $ for each subset of $\lbrace\ket{c_i}\rbrace$ due to the large number of subsets. To overcome this, we simplify our approach by considering a limited number of subsets. First, we limited ourselves to evolving QECCs with two-dimensional code spaces which leads to an $O\left(2^{2n-1}\right)$ number of subsets to consider for each tentative EC (a significant but not sufficient reduction). Second, by appealing to symmetry, we make the heuristic argument that we can fix one of the codewords, as $\ket{c_0}$ without loss, and only consider subsets of the form $\lbrace \ket{c_0}, \ket{c_i}\rbrace$ for $i = 1,\dots,2^{n}-1$. 


\section{Qubit overhead}\label{sec:qubit_overhead}

Let $n$ be the Hilbert space dimension of a QECC. Considering errors as a product of Pauli operators, we express a general error by assigning a Pauli letter for each entry of a vector of the form $\bm{E} = (a_1,a_2,\dots,a_n)$. Define $n_e(n,t)$ as the number of errors with weight $t$. For $t = 1$, each possible error can be constructed by choosing one of the $n$ entries of $\bm{E}$ and assigning one of three Pauli letters $\lbrace X,Y,Z\rbrace$. Therefore,
\begin{equation}
    n_e(n,1) = {n \choose 1} \times 3.
\end{equation}

\noindent For $t = 2$ the reasoning is similar: we choose two entries in $n$ to allocate the errors and, for each entry, we choose one of three Pauli letters. Thus,
\begin{equation}
    n_e(n,2) = {n \choose 2} \times 3^2.
\end{equation}

\noindent This reasoning holds true for any $t \leq n$. Therefore, the general formula for $n_e(n,t)$ is given by
\begin{equation}
    n_e(n,t) = {n \choose t} \times 3^t.
\end{equation}

For error-correction we are interested in detecting and correcting errors with weight up to $t$. Let $s(n,t)$ denote all possible errors up to weight $t$, i.e., the number of errors with weight less or equal to $t$. It follows that
\begin{equation}\label{eq:errorsuptot}
    s(n,t) = \sum_{i=1}^t n_e(n,i) =  \sum_{i=1}^t {n \choose i} \times 3^i.
\end{equation}

\noindent With $s(n,t)$ we can derive the minimum number of qubits necessary for constructing a non-degenerate quantum error-correction code capable of handling errors up to a weight $t$.

The argument goes as follows: if the codewords are made by $n$ qubits, then at most $n-1$ auxiliary ancilla qubits are employed in the syndrome measurement stage. Therefore, the syndrome is a vector with at most $n-1$ binary entries. Since  there exist $2^{n-1}$ binary vectors with $n-1$ entries and at least one distinct vector must be assigned to each particular error, there must exist at least as many binary vectors as the number of possible errors for a non-degenerate error correction code to be able to correct all errors up to weight $t$:
\begin{equation}\label{eq:ineq_overhead}
    s(n,t) + 1 \leq 2^{n-1}
\end{equation}

\noindent To account for the case in which no errors occurred, $1$ is added to the LHS of Equation \eqref{eq:ineq_overhead}. 

For $t = 1$, it follows
\begin{equation}\label{eq:overhead_t1}
    {n \choose 1} \times 3 + 1 = 3n + 1\leq 2^{n-1}.
\end{equation}

Note that $n = 5$ saturates the inequality \eqref{eq:overhead_t1}, therefore it is of no use to try to build a QECC with less then $5$ qubits; non-degenerate $5$-qubits codes that correct single-qubit errors are called perfect codes \cite{Gottesman1997,Laflamme1996} since they have the property of using every available syndrome for $5$ qubits.
\end{document}